\documentclass{nature}
\usepackage{times}
\usepackage{color}
\usepackage{amsmath, amssymb}
\usepackage{graphicx}

\makeatletter
\let\saved@includegraphics\includegraphics
\AtBeginDocument{\let\includegraphics\saved@includegraphics}
\renewenvironment*{figure}{\@float{figure}}{\end@float}
\makeatother
\newcommand{\onlinecite}[1]{\hspace{-1 ex} \nocite{#1}\citenum{#1}}

\title{Percolative Mott insulator-metal transition in doped Sr$_2$IrO$_4$}

\author{Zhixiang Sun$^{1,2,*}$, Jose M. Guevara$^2$, Steffen Sykora$^2$, Ekaterina M. P\"arschke$^3$, Jeroen van den Brink$^2$,
Kaustuv Manna$^{2,4}$, Andrey Maljuk$^2$, Sabine Wurmehl$^{2,5}$, Bernd B\"uchner$^{2,5,6}$, Christian Hess$^{2,6,\dagger}$}

\date{}
\begin{document}

\maketitle
\begin{affiliations}
 \item Center for Joint Quantum Studies $\&$ Department of Physics, Tianjin University, Tianjin, China
 \item IFW Dresden, Helmholtzstr. 20, 01069 Dresden, Germany
 \item Department of Physics, University of Alabama at Birmingham, Alabama 35294, USA
 \item Max-Planck-Institute for Chemical Physics of Solids, 01187 Dresden, Germany
 \item Institute for Solid State Physics, TU Dresden,  01069 Dresden, Germany
 \item Center for Transport and Devices, TU Dresden, 01069 Dresden, Germany
\end{affiliations}



\begin{abstract}
Despite many efforts to rationalize the strongly correlated electronic ground states in doped Mott insulators, the  nature of the doping induced insulator to metal transition is still a subject under intensive investigation. Here we probe the nanoscale electronic structure of the Mott insulator Sr$_2$IrO$_{4-\delta}$ with low-temperature scanning tunneling microscopy and find enhanced local density of states (LDOS) inside the Mott gap at the location of individual apical oxygen site defects. We visualize paths of enhanced conductance arising from the overlapping of defect states which induces finite LDOS at the Fermi level. By combining these findings with the typical spatial extension of isolated defects of about 2~nm, we show that the insulator to metal transition in Sr$_2$IrO$_{4-\delta}$ is of percolative nature.
\end{abstract}

\clearpage

Introducing defects in a Mott insulator leads to a fascinating appearance of different competing orders such as magnetism, superconductivity, or strange metallic behaviour which includes enigmatic collective ordering states with a characteristic pseudogap. Local probes as for example scanning tunneling microscopy have successfully been used in the past to investigate these phenomena, particularly in the case of the well-established cuprate superconductors\cite{Hanaguri2004,Kohsaka2007,SilvaNeto2014}. Recently, a new pathway
to studying Mott physics was found by the discovery that the spin-orbit coupling driven pseudospin $J_\mathrm{eff}=1/2$ Mott states of iridium oxide materials show an intriguing parallel to the spin $S = 1/2$ Mott phases of the cuprate materials\cite{kim2014fermi, cao2016hallmarks, kim2016observation}.

A prominent example is Sr$_2$IrO$_4$ which is a layered Mott-Hubbard insulator with a weak antiferromagnetic order below $T_\mathrm{N} \simeq 240$ K\cite{cao1998weak}, where the magnetic moments are canted\cite{Kim1329}. The material shares structural and magnetic similarities to the parent compounds of superconducting layered cuprates\cite{kim2012magnetic}.  Based on these observations, it has been argued that proper doping could drive the material into a high-temperature superconducting phase\cite{wang2011twisted}. Indeed, recent experiments have shown that with electron doping a pseudogap feature and an arc-like Fermi surface structure can be observed\cite{yan2015electron, kim2016observation, Battisti2017}. Furthermore, signatures of the spin-polaron physics well-known from underdoped La$_{2}$CuO$_{4}$ and typical for the presence of strong electron correlations\cite{Chernyshev2003} were recently found in Sr$_2$IrO$_4$ using tunneling spectroscopy\cite{Jose2019}.

One possible method to drive the Mott insulating system Sr$_2$IrO$_4$ into a metallic state is to substitute the elements Ir or Sr through chemical doping\cite{PhysRevB.86.125105, chen2015influence, de2015collapse}.
Recently, the possibility to introduce electrons by substituting La for Sr has gained particular interest because the insulator-metal transition caused by this doping scheme has been reported to be connected with the formation of a pseudogap, reminiscent of the pseudogap phase of the cuprates\cite{chen2015influence, de2015collapse}, including a short range stripe-like charge ordering pattern trapped by the La atoms observed by scanning tunneling microscopy
\cite{Battisti2017}.
However, in all the above studies the mechanism of the enhanced conductivity by doping remains unclear.

A further way to turn Sr$_2$IrO$_4$ into a metallic phase is to introduce dilute oxygen vacancies into Sr$_2$IrO$_{4-\delta}$ where transport probes have found an insulator-metal transition at $\delta \approx 0.04$\cite{korneta2010electron}. Similar to the effect of La doping, the removal of oxygen is also expected to result in electron doping.
Such oxygen depletion, like any other chemical doping scheme, also introduces disorder, which is expected to lead to localization of states around the defects which may become particularly strong in a quasi-2D system\cite{Mott1990}. Despite the expected localization caused by the impurities, it has been speculated, based on a variable range hopping analysis of resistivity data, that charge transport in oxygen deficient Sr$_2$IrO$_{4}$ is of percolative origin, where metallic patches are formed around the oxygen depleted sites\cite{korneta2010electron}.
But an experimental proof is lacking and thus the underlying microscopic mechanism for the emergence of metallicity remains to be clarified.

Here, we investigate intrinsic defects in oxygen deficient Sr$_2$IrO$_{4}$ using atomically resolved low temperature scanning tunneling microscopy/spectroscopy (STM/STS) and probe the
local electronic structure
with a focus on the apical oxygen site defects.
We observe characteristic bound states inside the Mott gap in a confined area around isolated defects, which are naturally explained by impurity scattering of the correlated electrons of the Mott state. Interestingly, in regions with a spatial overlap of several defect areas  a closing of the gap occurs.
Based on this result we conclude that the insulator-metal transition arises from the percolative creation of conductive paths within an antiferromagnetic background. This finding suggests that the metallic state of the electron doped Sr$_2$IrO$_4$ is distinctly different from the strange metal of underdoped cuprates where the common understanding is that its origin lies in the motion of doped charge carriers in a correlated antiferromagnetic background.

\begin{figure}[!t]
\begin{center}
\includegraphics[width=0.6 \textwidth]{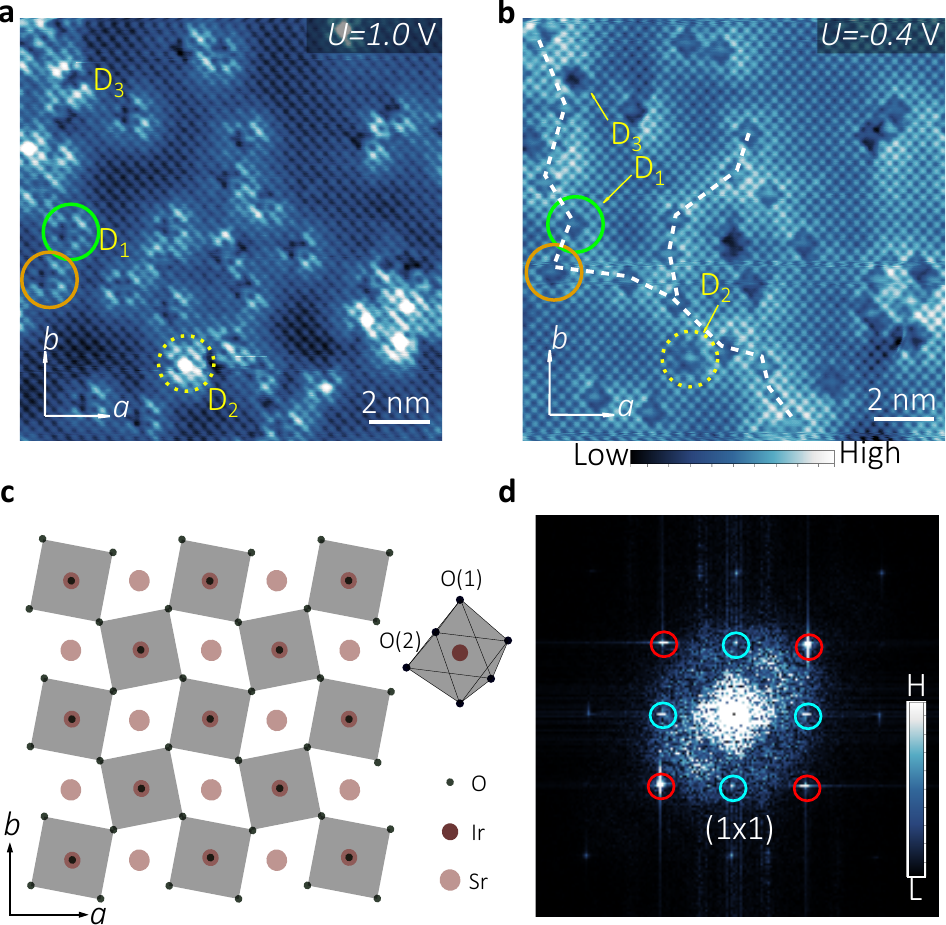}
\caption{\label{Fig:topo}}
\end{center}
\end{figure}

Figures \ref{Fig:topo}(a) and \ref{Fig:topo}(b) show a 14~nm $\times$ 14~nm field of view of atomically resolved topography data of a freshly cleave sample surface, measured at $T=9$~K using bias voltages $U = 1.0$~V and $U = -0.4$~V, respectively. Since the natural cleaving plane of Sr$_2$IrO$_4$ is in between the SrO double-layers, we attribute the visible corrugation to either the Sr or to the apical oxygen ions. Interestingly, the amplitude plot of the Fourier transform (Fig.~\ref{Fig:topo}(d)) reveals not only the Bragg peaks of a 2D square lattice but also a ($1\times1$) superstructure. It is consistent with a planar symmetry reduction which arises from an in-plane rotation of the IrO$_6$ octahedra (see Fig.~\ref{Fig:topo}(c)) by about 12$^\circ$ along the $c$ axis\cite{huang1994neutron}.

Let us first characterise the visible defects in the topographic data. A large-scale count (see Fig.~\ref{Fig:S1} in our Supplementary Information) yields a total abundance of about 2.6\% per Ir atom.
The defects can be differentiated by their distinct appearance at positive and negative bias voltages into three types denoted by D$_1$, D$_2$, and D$_3$, as is illustrated in Figs.~\ref{Fig:topo}(a) and \ref{Fig:topo}(b). The two types D$_1$ and D$_2$ are located at the same lattice site in the $ab$ plane but can be clearly distinguished from their height profiles in the topography taken at positive bias (see Fig.~\ref{Fig:topo}(a)).
Since our sample possesses a significantly reduced resistivity due to oxygen deficiency\cite{Jose2019}, it is reasonable to attribute this effect to the missing oxygen\cite{korneta2010electron}.
On this basis, we assign the visible regular atomic corrugation to the Sr atoms of the topmost layer, and the defects D$_1$ and D$_2$ to the apical oxygen (O(1) in Fig. \ref{Fig:topo}(c)) or the Ir positions.

A more careful inspection of the topographic data reveals that the intensity distribution in the environment of D$_1$ and D$_2$ has a certain chirality depending  on the bias voltage at which the topography is recorded (see Supplementary Information). This feature,
like the ($1\times1$) superstructure,
is naturally explained by the rotation of the IrO$_6$ octahedra and is well-known from previous STM measurements on the related double-layer perovskite Sr$_3$Ir$_2$O$_7$\cite{okada2013imaging}.

A third defect type (D$_3$) with a much lower surface abundance than D$_1$ and D$_2$ appears as a hole on the sample surface and is located at the surface site of the Sr ions. Unlike D$_1$ and D$_2$, which, as we will show further below, have a profound impact on the local density of states (LDOS), the effect of D$_3$ is rather weak. This indicates for D$_3$ either the possibility of a missing Sr atom or an isovalent impurity ion such as the light earth alkalines Ca or Mg.

From the topographic data in Figs.~\ref{Fig:topo}(a) and \ref{Fig:topo}(b) we clearly see an enhanced intensity within a few lattice constants around the defects. Consequently, since the topo\-graphy describes the energy-integrated LDOS we expect an appearance of new states near the Fermi level around the defects (bound states). These states may combine to conglomerates of conductive patches as indicated by dashed lines in Fig.~\ref{Fig:topo}(b) and offer the possibility of percolative charge transport. To confirm this expectation we have measured the tunneling spectrum on top of the discussed defect types and in a clean area of the sample (for experimental details see Supplementary information). The result is shown in Fig.~\ref{Fig: defect}. In the spectrum taken on a place free of defects (Fig.~\ref{Fig: defect}(a), the same as in Ref. \onlinecite{Jose2019}) we can clearly see the Mott gap $\Delta\approx620$~meV which is consistent with results from previous STM measurements\cite{dai2014local, yan2015electron}. Outside this gap there are features (indicated by red arrows) which are the signature of spin polaron physics\cite{Jose2019}. The impact of the defects is shown in Fig.~\ref{Fig: defect}(b). While on the D$_1$ and D$_2$ defect types $\Delta$ is reduced to a value of less than 150 meV, there is no significant decrease for the D$_3$ type. For the D$_1$ defect at both polarities
the spectral weight is clearly shifted towards the Fermi level, connected with peak-like features at about $-200$~meV and $250$~meV, resulting in a remaining gap of about 100 meV.
This transfer of spectral weight is seen more clearly in the difference of the spectra taken on defects and taken on the clean place as plotted in the lower panel of Fig. \ref{Fig: defect}(b). While for D$_1$ a clear resonance peak inside the gap is found, the filling of the gap with spectral weight is also present for D$_2$ but is more smooth and less resonant-like.

The appearance of the in-gap states can be confirmed by considering theoretically the effect of on-site potential scattering caused by local defects/charge. For this we employed the theoretical model describing spin-polarons in Sr$_2$IrO$_4$, which we introduced in Ref. Ref. \onlinecite{Jose2019} to describe the spectral signatures of a clean system. Here we additionally introduced a local defect and calculated its effect to the differential conductance on the basis of a t-matrix approach (see Methods for details). As shown in Fig.~\ref{Fig: theo}, characteristic bound states appear inside the Mott gap on top of a defect. As observed in our experiment, the defect  produces spectral weight near the Fermi level which appears additionally to the spin polaron excitations of the clean Mott system. If the distance of the individual defects is small enough these bound states combine and may form impurity bands which can contribute to charge transfer.

To underpin this picture, we experimentally investigate in Fig.~\ref{Fig: spec} the spatial extension of the defect bound states and show the variation of the tunneling spectrum crossing defects of type D$_1$ and D$_2$. Thereby we study the evolution of the spectra along two different directions in the sample (indicated by white and yellow arrows in Fig.~\ref{Fig: spec}(a)). In the first case (along the white arrow) the bound states of the affected defects are rather distinct. We can see that the effect is very localized at the defects and the Mott gap does not fully close on the individual defects, see Fig.~\ref{Fig: spec}(b). The spatial extension of the individual defect states is better seen in Fig.~\ref{Fig: spec}(c) where the line profile of the spectra is illustrated  using a pseudocolor scale. For D$_1$ and D$_2$ the bound states appear in areas of about 2 nm and 1.5 nm in diameter, respectively. The absence of states at the Fermi level (white dashed line in Fig.~\ref{Fig: spec}(c)) clearly indicates that the ground state is still insulating, even on top of the defects.

In the second line profile (along the yellow arrow in Fig.~\ref{Fig: spec}(a)) where we track the spectra starting from a clean region of the sample into an area with a large defect density the behaviour changes dramatically.
Within the area of high defect density
the gap nearly closes which indicates that we have entered a region where charge transport is possible. This observation indicates the possibility that the entered region is a part of a path along which percolative charge transfer can happen (see white dashed lines in Fig.~\ref{Fig:topo}(b)).

To answer the question at which doping such percolative transport is possible we have performed a standard bond percolation simulation (see Methods) for a 2D square lattice where we have fixed the size of the bond to 4 lattice constants (which corresponds to 2 nm extension of the impurity effect as found from Fig.~\ref{Fig: spec}). From this calculation we found a percolation threshold of 3.7\% defect concentration over the Ir atoms. Interestingly, this value is remarkably close to the critical doping of $\delta \approx 0.04$ where the insulator to metal transition is found in single-crystal Sr$_2$IrO$_{4-\delta}$\cite{korneta2010electron}.
This agreement of the percolation threshold with the critical doping level of the insulator-metal transition provides very strong evidence that the metallic state is realized by percolative conductive paths of overlapping impurity sites within an otherwise only weakly altered Mott insulating antiferromagnetic background. We stress that the metallic state has been reported to exhibit antiferromagnetic order in practically the same temperature range as pristine Sr$_2$IrO$_4$\cite{korneta2010electron}, which supports our scenario.

We mention that a similar closing of the Mott gap induced by oxygen vacancies has previously been seen by probing the differential conductance for the closely related higher order Ruddlesden-Popper iridate material Sr$_3$Ir$_2$O$_7$, albeit for a significantly smaller Mott gap ($\sim 130$~meV for the clean material)\cite{okada2013imaging}. The authors of this study argued that the origin of the gap filling could be a defect-induced reduction of the on-site Coulomb repulsion. According to band structure calculations this effect would cause an almost rigid shift of the respective two parts of density of states towards each other which are related to the lower and upper Hubbard band edges. Such a scenario is, however, incompatible with our data since the spectral weight which is transferred into the gap at a single impurity clearly appears as distinct peaks, characteristic of bound states, at both edges of the gap rather than a shift of the band edges (see Fig.~\ref{Fig: defect} b)). We therefore conclude that the new in-gap states indeed are the result of impurity scattering, as calculated in Fig.~\ref{Fig: theo}.

It is further interesting to note that the $\mathrm{d}I/\mathrm{d}U$ at high impurity density as shown in Fig.~\ref{Fig: spec}(d) strongly resembles that of the reported pseudogap in La-doped Sr$_2$IrO$_4$\cite{Battisti2017}. Indeed, in such a doping scheme a similar scenario for the insulator-metal transition arising from overlapping impurity scattering bound states might apply. In this case the percolative conduction paths might be caused by the La$^{3+}$-ion impurity potential. While a careful on-site defect spectroscopy, as we did here for the first time for oxygen defects, should be able to unequivocally clarify such a notion,
the importance of the impurity scattering strength for the formation of states inside the Mott gap is underpinned by a recent study of La-doped Sr$_3$Ir$_2$O$_7$\cite{Wang2019}. Those results clearly show that the impact of the dopants on the electronic structure decisively depends on their lattice position and thus the strength of the impurity potential within the IrO$_2$ layers.

To conclude, in this work we have investigated the effect of intrinsic oxygen vacancy defects in Sr$_2$IrO$_4$ to the Mott insulating state and its local electronic structure. We have shown that characteristic bound states are formed within a spatial scale of around 2 nm,
creating spectral weight inside the Mott gap. For isolated defects the gap does not fully close but when defects start to spatially overlap the gap closing becomes almost complete.  Therefore, our local probe clearly reveals the scenario of filamentary metallic-like conductivity through agglomerations of defects.
This interpretation is fundamentally different from the usual mechanism of the insulator to metal transition in the underdoped cuprates which usually is understood as to originate from the correlated motion of charge in an antiferromagnetic background.
Our finding is corroborated by a simulation of the bond percolation threshold based on realistic parameters of the lateral extension of the impurity states, where we find a very good agreement with the experimental critical doping level.

\begin{methods}

\section{Experimental details}
Sr$_2$IrO$_{4-\delta}$ single crystals with reduced resistivity due to oxygen deficiency were grown by the flux method\cite{Jose2019}.
The STM measurements were performed in a home-built variable temperature STM\cite{schlegel2014design}, using a PtIr tip. The single crystals were cleaved inside the STM head in cryogenic vacuum at about 9 K with the $c$-axis as the surface normal.
For the measurements of the tunneling conductance spectra we applied the standard lock-in method with an external modulation of 1.1111 kHz. The root mean square of the alternating voltage modulation that we applied for acquisition the tunnelling spectra is 20 mV.

\section{Numerical simulation}

\subsection{\textit{Simulation of the differential tunneling spectrum}}

The LDOS calculation has been performed on the basis of a standard t-matrix approach where the momentum integrated Green's function of the clean system is taken from Ref. \onlinecite{parschke2017correlation}.

We consider the one-particle Green's function $G_0({\bf k},E)$ as obtained from the self-consistent Born approximation plus a local potential scatterer with an assumed potential $V_{{\bf k},{\bf k}'} = V_{imp}$ in momentum space. These parameters are used for a standard t-matrix approach to calculate the Fourier-transformed local density of states,
\begin{eqnarray}
\rho ({\bf q},E) = \frac{1}{\pi} \sum_{\bf k} \Im  G({\bf k},{\bf k} - {\bf q},E),
\label{LDOS}
\end{eqnarray}
which is the quantity that can be directly measured by momentum-resolved STM/STS experiments. $G({\bf k},{\bf k}',E)$ is the retarded Green's function in the presence of one single impurity and is related to the retarded Green's function $G_0({\bf k},E)$ of the bulk material via the equation
\begin{eqnarray}
G({\bf k},{\bf k}',E) = G_0({\bf k},E) + G_0({\bf k},E) T_{{\bf k},{\bf k}'}(E)G_0({\bf k}',E),
\label{G_imp}
\end{eqnarray}
where the energy-dependent t-matrix $T_{{\bf k},{\bf k}'}(E)$ is determined by the following self-consistency equation,
\begin{eqnarray}
T_{{\bf k},{\bf k}'}(E) = V_{{\bf k},{\bf k}'} + \sum_{{\bf k}''} V_{{\bf k},{\bf k}''} G_0({\bf k}'',E)T_{{\bf k}'',{\bf k}'}(E).
\label{t-matrix}
\end{eqnarray}
Using  the two functions $G_0({\bf k},E)$ and $V_{{\bf k},{\bf k}'}$ as input parameters we have solved Eq.~\eqref{t-matrix} self-consistently. The obtained result for $T_{{\bf k},{\bf k}'}(E)$ is inserted in Eq.~\eqref{G_imp} for the full Green's function $G({\bf k},{\bf k}',E)$. Finally, this result is used in Eq.~\eqref{LDOS} to evaluate the momentum-dependent variation of the density of states $\rho ({\bf q},E)$ due to the presence of the impurity.

The corresponding impurity-induced variation of the density of states in real space is given by the Fourier transform of Eq.~\eqref{LDOS},
\begin{eqnarray}
\rho({\bf R}_i,E) = \frac{1}{N} \sum_{\bf q} \rho ({\bf q},E) e^{i{\bf q}{\bf R}_i},
\end{eqnarray}
where ${\bf R}_i$ is a lattice vector and $N$ is the total number of lattice sites in the system. Note that this quantity is proportional to the experimentally probed differential conductance $dI/dU$. The result for $\rho({\bf R}_i = 0,E)$ as a function of $E=U$ is shown in Fig.~\ref{Fig: theo} (yellow line). Here the specific location of the impurity, i.e. ${\bf R}_i = 0$ is considered. For comparison, the calculated spectrum without impurity is also shown (green line).

\subsection{\textit{Bond percolation analysis}}

The numerical simulation of the bond percolation\cite{Newman2000,Sykes1964} was performed on a 2D square lattice of $100 \times 100$ lattice sites according to the following procedure. At first we fixed the total number of defects to some chosen value and generated in total 1000 random distributions of these defects. For each of these distributions we counted the number of unbroken  paths of connected defects which link at least two opposite boundaries of the lattice. If we found at least one of such paths for each single distribution we assigned  the corresponding doping level metallic character.

Two defects were defined as connected (bond) if charge transfer is possible between these defects via the emergence of at least one bound state which is lower in energy than the states of a single defect. Experimentally, this situation appears in Fig.~4(d) where the Mott gap is closed between two distinct defects. From these considerations we fixed for our percolation simulation the size of a bond to 4 lattice constants.

The above steps were repeated for different doping levels and for each doping value we assigned either insulating or metallic character according to the criterion defined above. We found metallic character above a percolation threshold of $3.7\%$ defect concentration.

\end{methods}



\begin{addendum}
  \item [Acknowledgements]
    We thank C. Renner, B. J. Kim, V. Madvahan and M. Allan for helpful discussions and D. Baumann and U. Nitzsche for technical assistance. The project is supported by the Deutsche Forschungsgemeinschaft through SFB 1143 and by the Emmy Noether programme (S.W. Project No. WU595/3-3). Furthermore, this project has received funding from the European Research Council (ERC) under the European Unions' Horizon 2020 research and innovation programme (Grant Agreement No. 647276-MARS-ERC-2014-CoG).

  \item[Autor contributions]
    K.M., A.M., and S.W. grew the single crystals, Z.S. and J.M.G. carried out the STM experiments. S.S., E.M.P., and J.v.d.B. performed the theoretical calculations. Z.S., S.S. and C.H. wrote the manuscript. C.H. and B.B. designed the study. All authors contributed to the discussion of the results and the manuscript.

  \item[Competing financial interests]
  The authors declare no competing financial interests.

  \item[Materials \& Correspondence]
  To whom correspondence should be addressed; E-mail: c.hess@ifw-dresden.de

\end{addendum}

\clearpage

\section*{Figure Captions}

\noindent\textbf{Fig.~1. Surface topography and the different types of defects.} (\textbf{a, b}) Topography of the sample surface measured at $U = 1.0$~V and $U = -0.4$~V, respectively. Characteristic defects are marked by circles. The white dashed lines indicate possible paths of percolative charge transport. (\textbf{c}) A top view of the lattice structure in the $ab$-plane of Sr$_2$IrO$_4$. In the right side octahedron, the vertical direction is along $c$ axis. O(1) is the apical oxygen and O(2) is the basal oxygen. (\textbf{d}) The amplitude of the Fourier transform of the topography image (b), where we label the Bragg peaks and the peaks due to the ($1\times1$) lattice distortion. \\

\noindent \textbf{Fig.~2. Differential tunneling conductance spectra.}  (\textbf{a}) Overview tunnelling conductance spectrum taken at a clean place. (\textbf{b}) Spectra taken on top of the D$_1$, D$_2$, and D$_3$ defects as discussed in the text. Lower panel: The differences between the spectra taken on the clean places and on top of the defects. The inset illustrates the transfer of the spectral weight into the gap on top of the apical oxygen site defects.
\\

\noindent\textbf{Fig.~3. Calculated effect of a local defect to the differential conductance.} The spectrum (yellow line) is characterized by an on-site Coulomb scattering potential with a strength of $V_{imp} = 0.028$ eV. The corresponding  differential conductance calculated for the clean system is also shown (green line).\\

\noindent\textbf{Fig.~4. The spatial electronic effects of different defects.}  (\textbf{a}) The corresponding topography where we have taken the spectroscopic mapping. Setpoint: $U= 1.0$~V, $I = 200$~pA. (\textbf{b}) The line profile of the spectra along the white arrow as shown in \textbf{a} crossing the two different types of defect, D$_1$ and D$_2$, shown with a constant offset for clarity. (\textbf{c}) The corresponding pseudocolor spatial map of the spectral intensity profile shown in \textbf{b}. (\textbf{d}) Line profile of the differential tunneling conductance spectra along the yellow arrow shown in \textbf{a}.

\clearpage

\clearpage

\begin{figure}[!t]
\begin{center}
\includegraphics[width=0.5\textwidth]{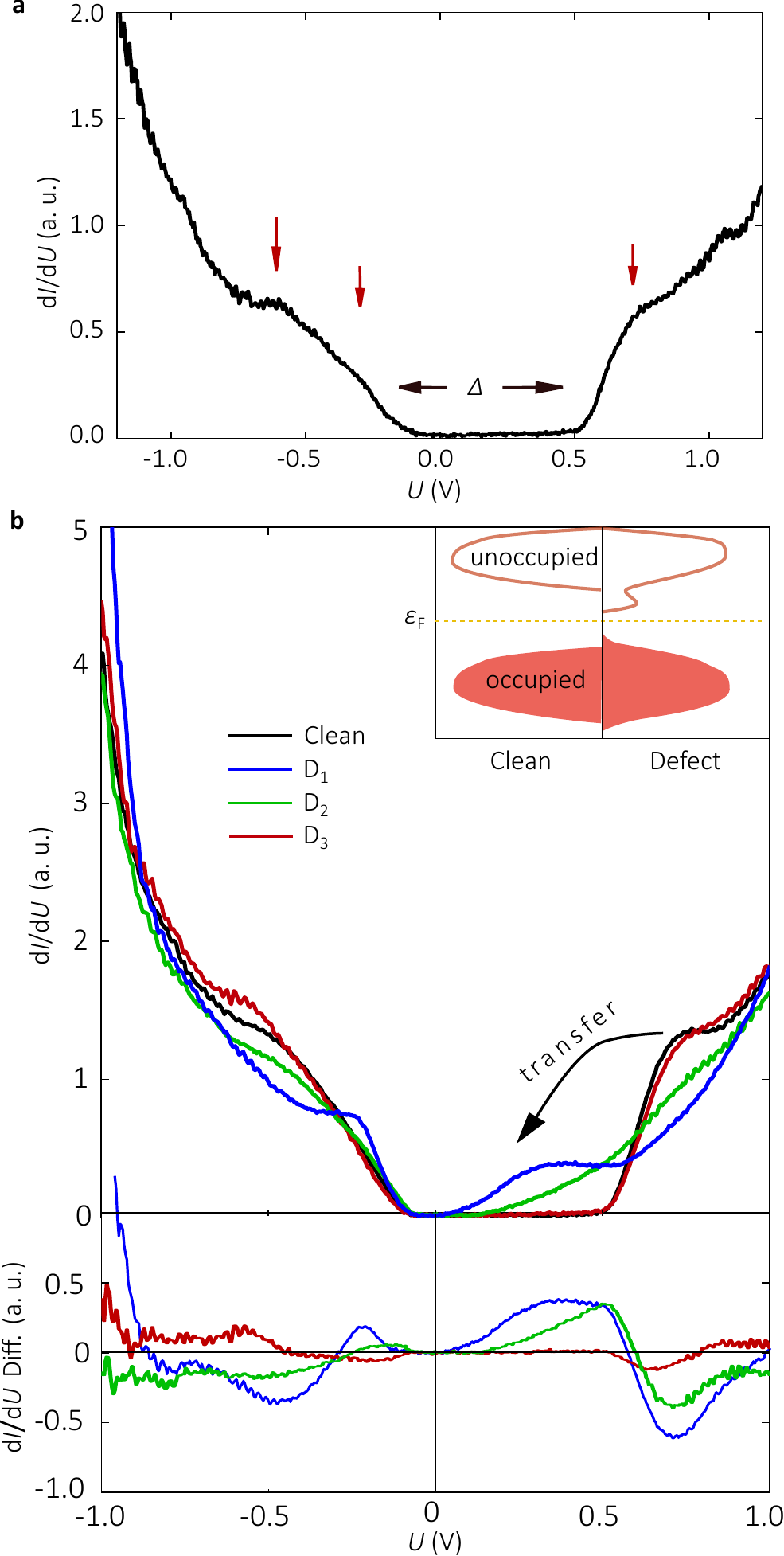}
\caption{\label{Fig: defect}}
\end{center}
\end{figure}

\clearpage

\begin{figure}[!t]
\begin{center}
\includegraphics[width=0.50\textwidth]{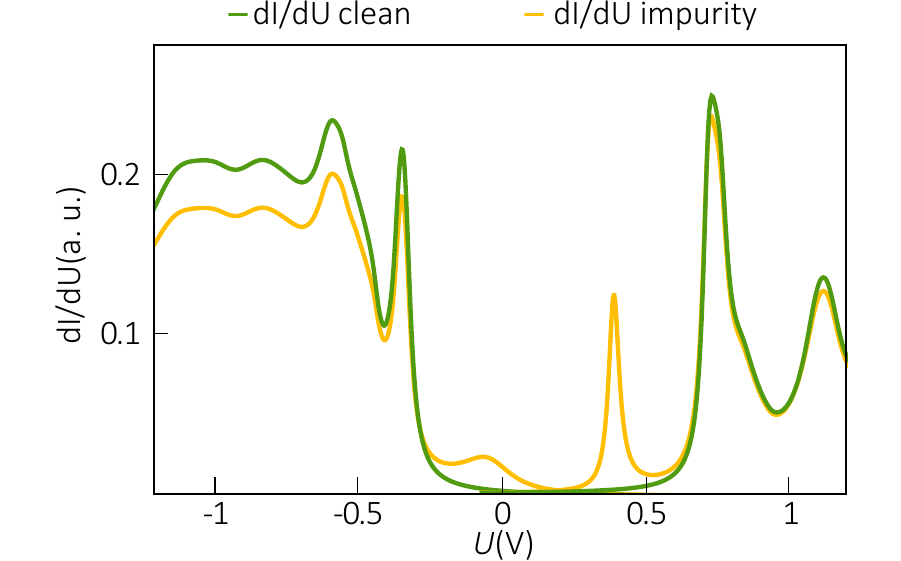}
\caption{\label{Fig: theo}}
\end{center}
\end{figure}

\clearpage

\begin{figure}[!t]
\begin{center}
\includegraphics[width=\textwidth]{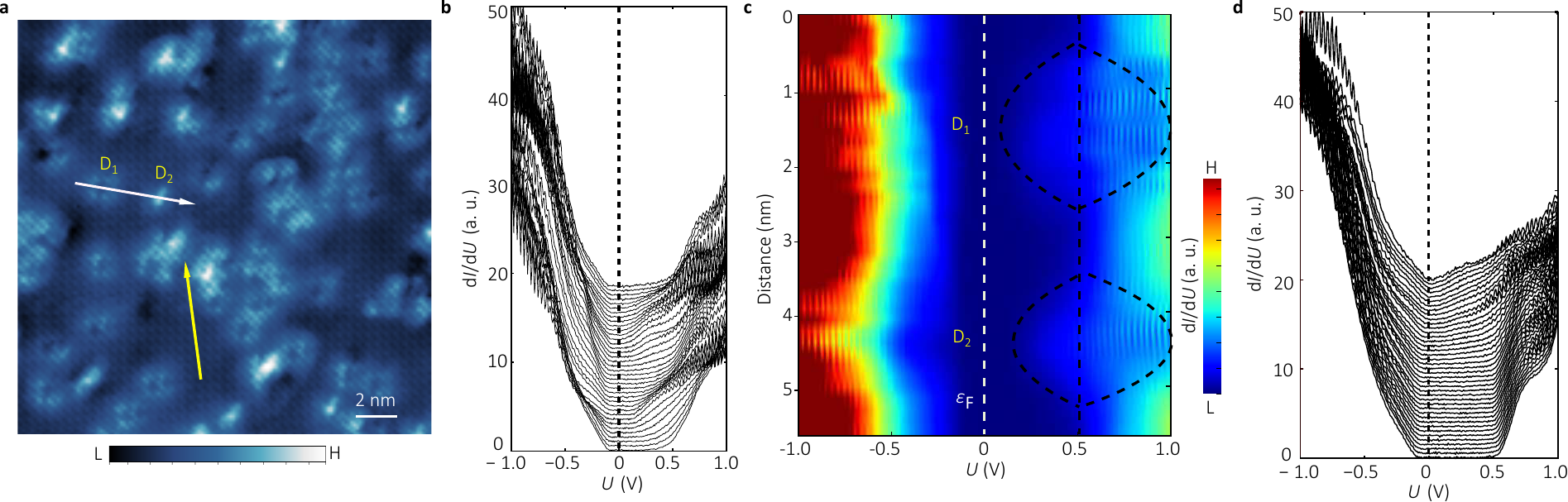}
\caption{\label{Fig: spec}}
\end{center}
\end{figure}


\clearpage
\renewcommand{\thepage}{\arabic{page}} \setcounter{page}{1}

\begin{center}
 \LARGE Supplementary Information for \vspace{24pt}

 \large\textbf{Percolative Mott insulator-metal transition in doped Sr$_2$IrO$_4$}
 \vspace{24pt}

 \normalsize
\baselineskip12pt
 Zhixiang Sun, Jose M. Guevara, Steffen Sykora, Ekaterina M. P\"arschke, Jeroen van den Brink,
Kaustuv Manna, Andrey Maljuk, Sabine Wurmehl, Bernd B\"uchner, Christian Hess

 \vspace{12pt}
 correspondence to c.hess@ifw-dresden.de
\end{center}
\vspace{12pt}

\baselineskip12pt
\noindent\textbf{This PDF file includes:}\\
\par
\begingroup
\leftskip=1.5cm 
\noindent
Fig.~\ref{Fig:S1} to \ref{Fig:S2}\\
\par
\endgroup

\clearpage

\renewcommand{\thefigure}{S\arabic{figure}}
\setcounter{figure}{0}

\section*{Defect density}

\begin{figure}[h]
\centering
\includegraphics[width=0.5 \textwidth]{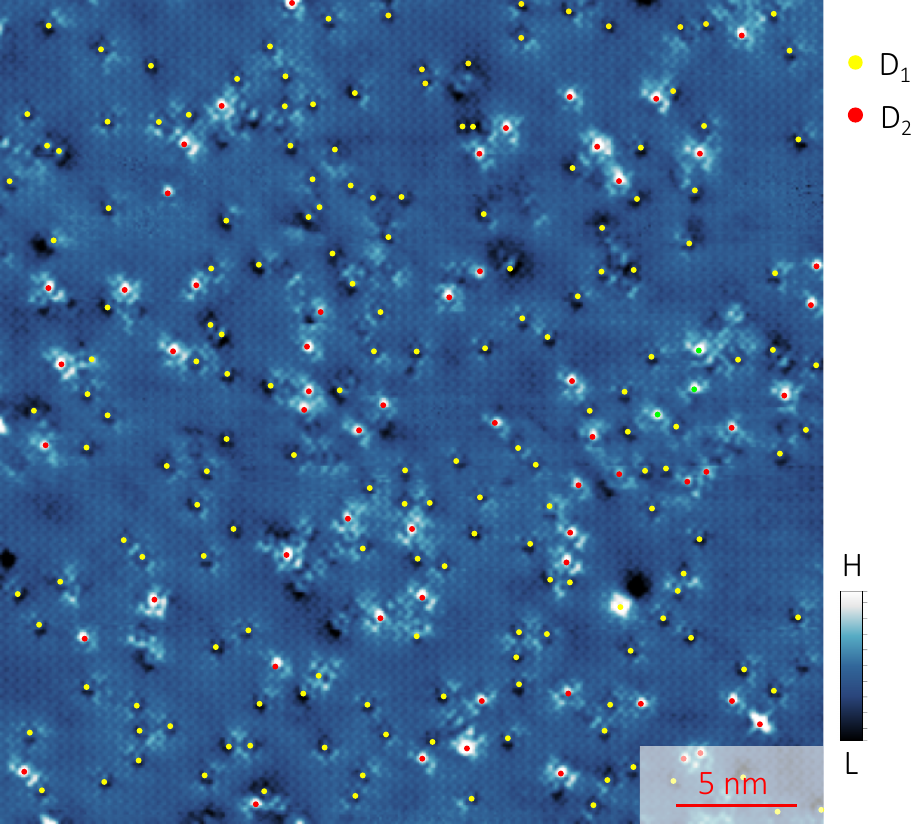}
 \caption{\label{Fig:S1} \textbf{Surface defect density counting.} A large area topography of the sample surface, taken with the parameter values $U = 1.0$ V, $I = 200$ pA. From the image we can see that the intrinsic defects are rather homogeneous. We have marked out two types of defects, D$_1$ (yellow) and D$_2$ (red).}
\end{figure}

We noticed that the D$_1$ and D$_2$ defects have different surface densities. One of the large area topographies is shown in Fig.~\ref{Fig:S1}. We can see that the D$_1$ defects have a larger surface density, about 2\% over the Ir atoms. The surface density of the D$_2$ defects is about 0.6\% over the Ir atoms.

\section*{Chiral feature in the topography}

\begin{figure*}[h]
\centering
\includegraphics[width=0.7\textwidth]{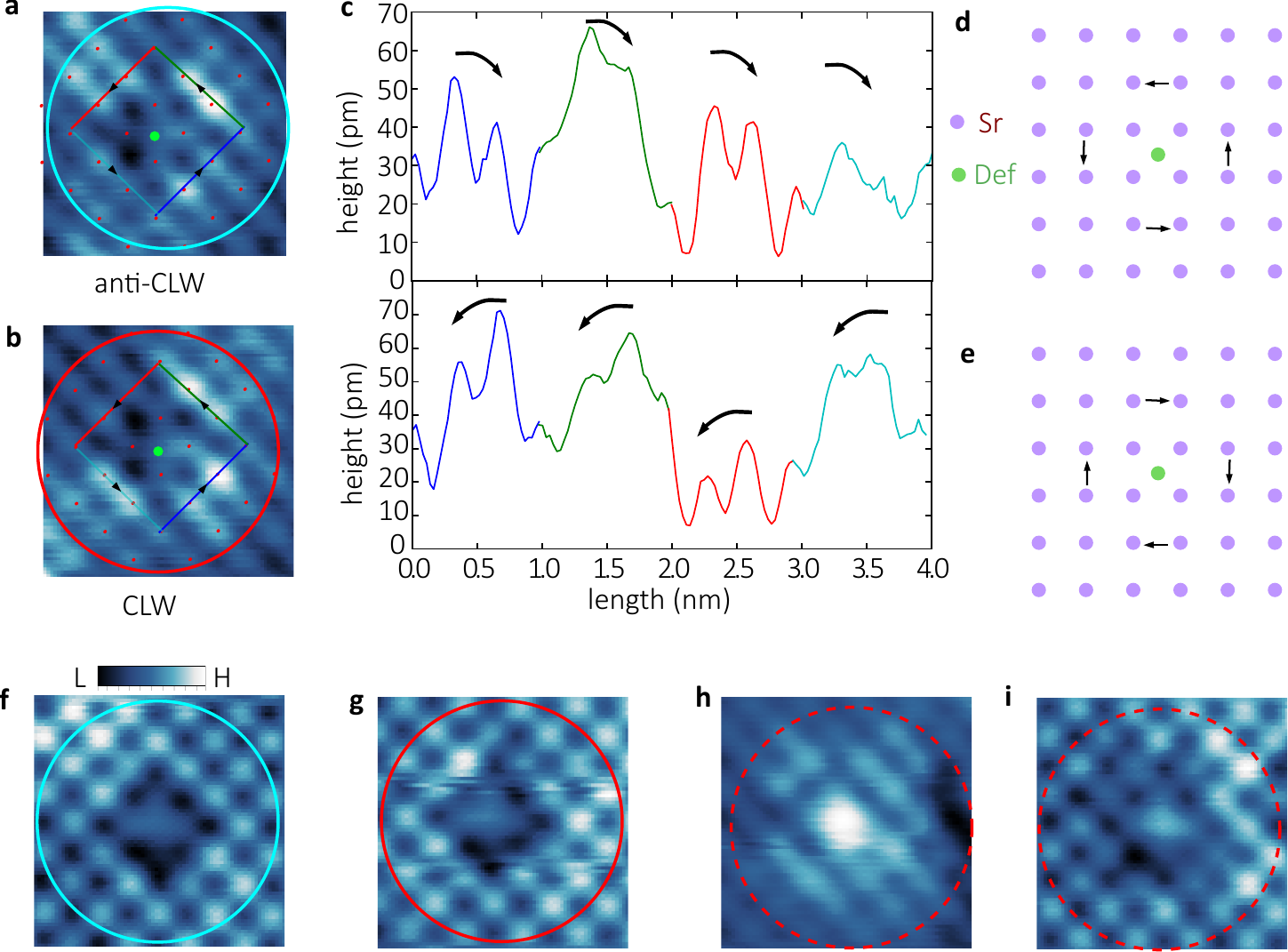}
 \caption{\label{Fig: chr} \textbf{Chiral behavior of the defect topographies and their bias dependencies.} The defect topographies are taken from Fig.~1. (\textbf{a},\textbf{b}). The anticlockwise and clockwise behavior around the D$_1$ type defects, which are taken from Fig.~1(\textbf{a}). The red dots are the positions of the surface Sr atom. (\textbf{c}) The line profiles of the topography marked in \textbf{a} and \textbf{b}, where the arrows are the guide to eyes on the change of the height profile. (\textbf{d},\textbf{e}) The illustrations of the D$_1$ defects topography images chirality shown in panels \textbf{a} and \textbf{b}, respectively. (\textbf{f},\textbf{g}) The topographies of the two defects in (\textbf{a},\textbf{b}) taken with a negative bias. (\textbf{h}) A D$_2$ type defect topography taken with a positive bias. (\textbf{i}) The D$_2$ type defect (shown in \textbf{h}) topography taken with a negative bias.}
\end{figure*}

\begin{figure}[h]
\centering
\includegraphics[width=0.5 \textwidth]{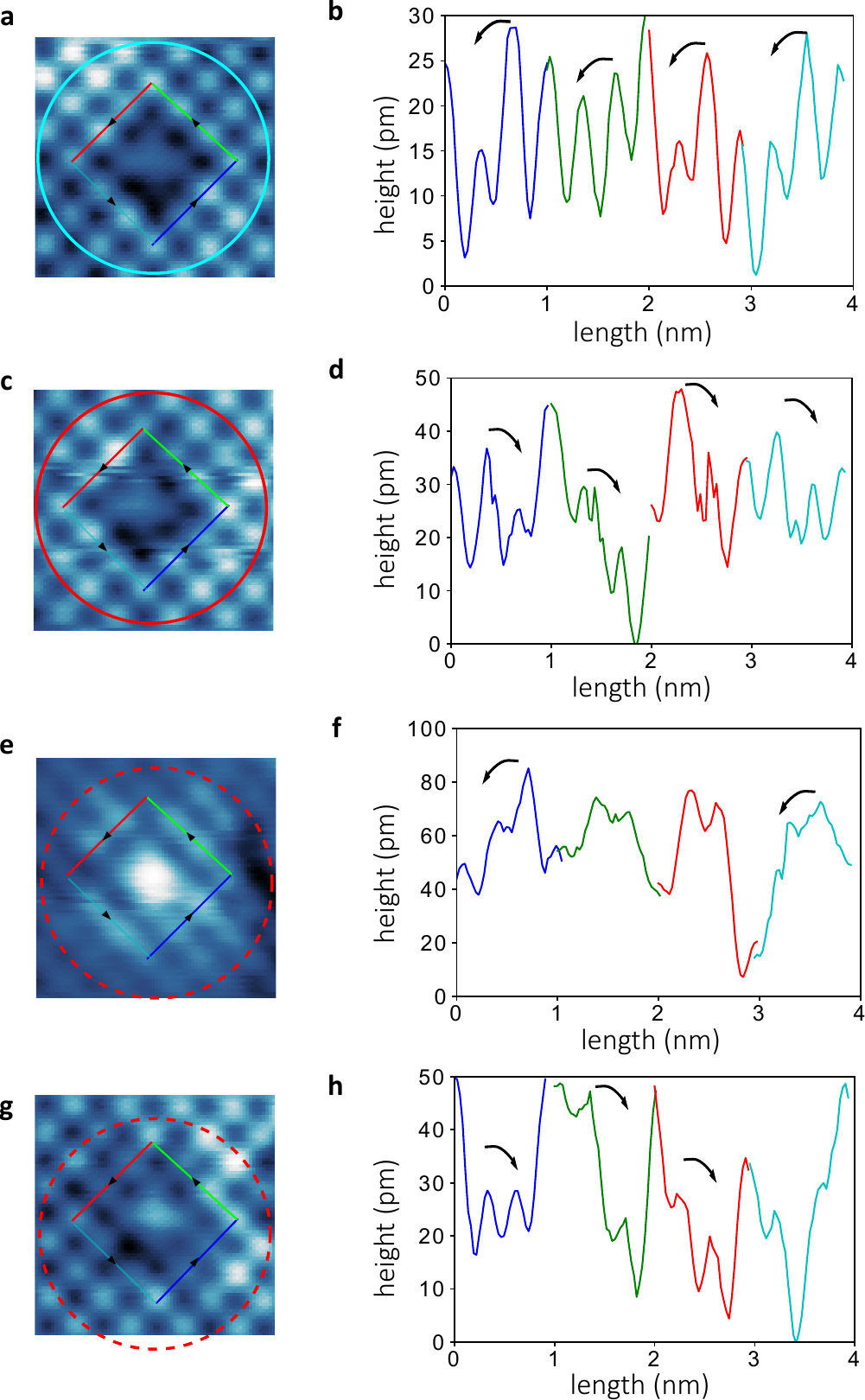}
 \caption{\label{Fig:S3} \textbf{More topographic height line profiles around D$_1$ and D$_2$ type defects shown in Fig. 1.} (\textbf{a}), (\textbf{c}), (\textbf{e}), (\textbf{g}) Topographies of individual defects. (\textbf{b}), (\textbf{d}), (\textbf{f}), (\textbf{h}) The corresponding line profiles around the defects.}
\end{figure}

A more careful comparison of the topographic appearance of the defects D$_1$ and D$_2$ reveals that each of these defect types possesses two variants which differ in the chirality of the surrounding topography.
This can be inferred from the close-ups of the topography of two different  D$_1$ defects as shown in Fig.~\ref{Fig: chr} (a), (b). While the four nearest surface atoms around the defect are hard to resolve, the respective height profiles of the next nearest Sr atoms represent the signature of the different chirality quite clearly. This signature is emphasized in the respective height profiles of the next nearest Sr atoms around the defects (panel (c)) and the illustrated chiralities (panels (d) and (e) of Fig.~\ref{Fig: chr}).
This topographic chiral character is also visible at negative bias voltage as can be inferred from the panels (f) and (g) of Figure~\ref{Fig: chr}. Apparently, the chiral signature in the height profile is opposite from that identified at positive bias (see Fig.~\ref{Fig: chr}).
A similar chiral dichotomy is also present for the D$_2$ defects (see panels  (h) and (i) of Figure~\ref{Fig: chr}). Their height profiles can been seen in Fig.~\ref{Fig:S3}.

Since D$_1$ and D$_2$ defects can be identified with the apical oxygen/Ir lattice sites, the chiral feature around these defects is naturally explained by the known rotation of the IrO$_6$ octahedra.
We mention that a chiral structure around defects has also been observed in previous STM measurement on the related double-layer perovskite Sr$_3$Ir$_2$O$_7$\cite{okada2013imaging}, which possesses a similar in-plane lattice distortion.

A third defect type (D$_3$) with a much lower surface abundance than D$_1$ and D$_2$ appears as a hole on the sample surface and is located at the surface site of the Sr ions. It has only a very weak effect on the local density of states indicating a minor impact on the charge environment. Therefore, besides of a missing Sr atom another possible assignment to D$_3$ would be an isovalent impurity ion such as the light earth alkalines Ca or Mg.

Beside of the chirality, we also notice that from the height profiles around the defect, the $C_4$ lattice symmetry in the $ab$ plane is clearly broken.
There have been many discussions about the structural symmetry breaking in Sr$_2$IrO$_4$. By carefully checking the topography we notice that the C$_4$ symmetry is fully broken around the D$_1$ and D$_2$ defects (see e.g.~Fig.~\ref{Fig: chr}(a,b)). This observation seems to indicate that there is more symmetry breaking than the equal rotation of the nearest IrO$_6$ octahedra, even considering the magnetic ordering at low temperatures (C$_2$ symmetry), e.g., the symmetry breaking of the rotation angle or the lattice size of nearest IrO$_6$ octahedra\cite{torchinsky2015structural}. This indicates that the structure of the material is more complicated than we expected and need to be further clarified.

\begin{figure}[h]
\centering
\includegraphics[width=0.5 \textwidth]{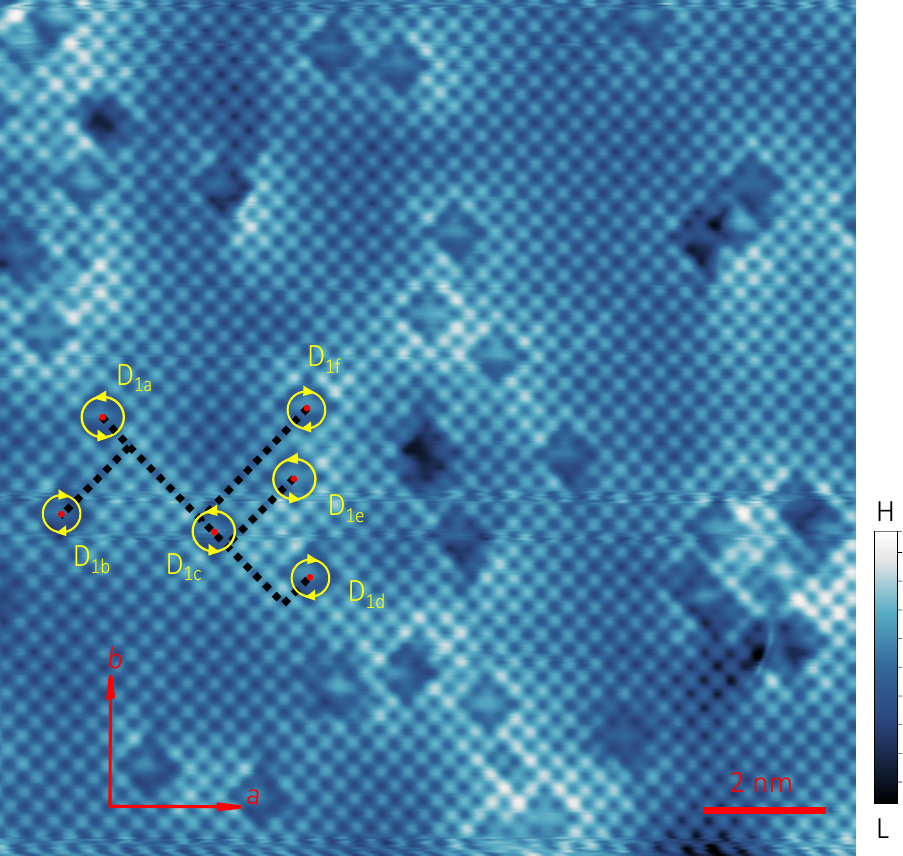}
 \caption{\label{Fig:S2} \textbf{The chiral behavior of the defect as a function of lattice site.} The topography is the same as the one shown in Fig. 1\textbf{a}. Yellow circles indicate the chirality of the corresponding D$_1$ type defects.}
\end{figure}

Let us first define the distance $R$ between two defects as the steps needed to get from one to the other by moving along the nearest Ir-Ir lattice directions in unit of $a_0$, where $a_0$ is the distance between two nearest Ir atoms. Then the Mod$(R, 2a_0)$ should be either $0$ or $a_0$, where Mod is the modulo operation. According to our observation, defects obeying Mod$(R,2a_0)=0$ have the same local chirality. In Fig.~\ref{Fig:S2}, we have marked out six D$_1$ type defects as D$_{1i}$ ($i=a,\dots,f$). Then we can apply the modulo computation to the distances between any two defects, but we consider here only the distance to D$_{1a}$:
\begin{eqnarray*}
\mbox{Mod}(R(\mbox{D}_{1b},\mbox{D}_{1a}), 2a_0) &=& \mbox{Mod}(7a_0, 2a_0) = 1a_0 \\
\mbox{Mod}(R(\mbox{D}_{1c},\mbox{D}_{1a}), 2a_0) &=& \mbox{Mod}(8a_0, 2a_0) = 0 \\
\mbox{Mod}(R(\mbox{D}_{1d},\mbox{D}_{1a}), 2a_0) &=& \mbox{Mod}(15a_0, 2a_0) = 1a_0 \\
\mbox{Mod}(R(\mbox{D}_{1e},\mbox{D}_{1a}), 2a_0) &=& \mbox{Mod}(14a_0, 2a_0) = 0 \\
\mbox{Mod}(R(\mbox{D}_{1f},\mbox{D}_{1a}), 2a_0) &=& \mbox{Mod}(15a_0, 2a_0) = 1a_0
\end{eqnarray*}
From these numbers, we can expect that D$_{1c}$ and D$_{1e}$ have the same chirality as D$_{1a}$ and the remaining defects have the opposite chirality, as is also observed in Fig. \ref{Fig:S2}.

\clearpage

%
%
%

\end{document}